\documentclass[preprint]{aastex}

\begin{document}

\title{Interpretations of the Accelerating Universe}

\author{J.V. Narlikar\altaffilmark{1}, R.G. Vishwakarma\altaffilmark{1} and G. Burbidge\altaffilmark{2}}

\altaffiltext{1}{Inter-University Centre for Astronomy and Astrophysics, Post bag 4, Pune 411007, India}
\altaffiltext{2} {Center for Astrophysics and Space Sciences,
  University of California 0424, San Diego,
  CA 92093-0424, USA}
\medskip

\begin{abstract}
It is generally argued that present cosmological observations support the accelerating models of the universe, as driven by a cosmological constant or `dark energy'.  We argue here that an alternative model of the universe is possible, which explains the current observations of the universe. We demonstrate this with a reinterpretation of the magnitude-redshift relation for Type Ia supernovae, since this was the test that gave a spurt to the current trend in favour of the cosmological constant.
\end{abstract}

\bigskip

\section{Introduction}

There are only two instances in modern cosmology in which a genuine 
theoretical prediction has been directly confirmed by observation.  
The first of these was the theoretical prediction obtained directly
from the solution of Einstein's equations by Friedmann (1922) and 
later Lemaitre (1927) that the universe is expanding (Hubble 1929).
The second was the theoretical prediction made in the framework of
the original steady-state cosmology (Bondi and Gold 1948; Hoyle 1948)
that the universe should be accelerating.  This has
been demonstrated in modern times by Perlmutter et al (1999) and by
Riess et al (1998, 2000).  

Unfortunately while the latter was a clear cut theoretical prediction,
the observers almost unanimously have chosen to interpret their 
detection of acceleration by maintaining the (standard) and highly
popular big bang model (SBBC) explaining the acceleration in terms of a non-zero
cosmological constant.  Because of this lack of balance in the
discussion and the interpretation of the acceleration, we finally published
a paper (Banerjee et al 2000) showing that the observational data available
up to that time can very reasonably be interpreted in terms of the 
quasi-steady state cosmological (QSSC) model.  In this model we had shown
that the QSSC model provides a good fit to the above data provided one takes
account of the absorption of light by intergalactic dust in the form of 
metallic whiskers.  Dust of this kind is required by the QSSC model to explain
the cosmic microwave background (Hoyle, Burbidge and Narlikar 1994$a$, 2000).  

Since that paper was published, new data have been obtained (Riess et al 2001)
involving a type Ia supernova at a much higher redshift.

In order to give some level of balance to the claim that is now being 
associated with these discoveries (cf. Flippenko 2001), in this paper we
extend our earlier discussion to see how well (or how badly) the  quasi-steady state model fits
the new data.  However, we first begin with a short discussion 
of why the cosmological constant currently enjoys so much popularity. 

\section{The advantages of $\lambda$}

The recent popularity of the cosmological constant with the proponents of standard
cosmology rests on the following aspects :\\

\noindent 1.  The constraints of `age' are relaxed.  By choosing a $\lambda$ suitably
it is possible to increase the age of the universe.  Thus the age of a standard flat
$\lambda = 0$ model is $\sim 6.5 \times 10^{9}h_{0}^{-1}$ yrs, for $H_{0}=100~h_{0}{\rm ~km ~s}^{-1}{\rm ~Mpc}^{-1}$, which is too low for accommodating globular cluster ages
of $\sim 11-13 \times 10^{9}$ yrs, for $h_{0}\approx 0.7$.  By way of comparison, the age
of a flat model with $\Lambda_0 \equiv \lambda/3H_{0}^{2} =0.7$  is
$\approx 13.5 \times 10^{9}$ yrs, for $h_{0}=0.7$.   At the same time it should be stressed that
the choice $h_{0} = 0.7$ is rather arbitrary.  While there are those who strongly advocate the
model with this value, there is compelling evidence from Saha et al. (2001)
that $h_{0} \lesssim 0.6$.

\medskip

\noindent 2.  The use of $\lambda$ allows the metric distance to any given redshift to be larger than the corresponding distance (for the same redshift) for the $\lambda=0$
model.  This was demonstrated very clearly by the magnitude-redshift relation
for Type Ia supernovae.  The excess faintness of supernovae (over the 
predictions of the $\lambda = 0$ Friedmann models) was interpreted as the 
proof of a non-zero, positive $\lambda $.  

\medskip

\noindent 3.  Indirect checks on $\lambda \neq 0$ are provided by the
structure formation scenarios and the observed inhomogeneities of the 
cosmic microwave background.  Here the choice of a non-zero $\lambda$ 
helps by expanding the parameter space and thereby relieving pressures on 
structure formation scenarios from the various observational constraints.  For a review
of these and other observational constraints see Bagla eta al (1996) and Narlikar
and Padmanabhan (2001). 

\medskip

\noindent 4.  If one insists on a flat model ($-$ as is implied by most kinds of inflation$-$) , the
observed baryonic density of matter is only a few percent of the total 
cosmological density.  In that case the rest ($\gtrsim$ 90\%) of all matter
has to be dark and probably non-baryonic.  The requirement of locating and
identifying so much of matter in this dark and esoteric form is not so severe
if we assume that a substantial part of it is taken over by $\lambda$ and is
called `dark energy' or `quintessence'.  Although the problem of bridging
the gap between the predicted and the observed still remains, the gap is
now to be explained more easily by an `intangible' dark energy than by a
more tangible dark matter particle.  

\bigskip

\section{The disadvantages of $\lambda$}

The gifts brought by the cosmological constant come accompanied by a few unwelcome
effects too.   We name a few below.

\medskip

\noindent 1.  With the popularity of the inflationary model, the cosmological constant found a
new life with its claimed origin in the force generated by transitions of the vacuum.  It is this force that is supposed to drive the universe in a transient de Sitter-like mode. It is tempting to suppose that the present $\lambda$ owes its origin to the inflationary phase. However,
Weinberg (1989) has commented on the extremely low value of $\lambda$
as required by the present cosmological observations compared to the primordial scenarios
that drove the universe into an inflationary mode.  If the inflation took place at the GUT 
epoch, the present value of $\lambda$ is too low by a factor $\sim 10^{-108}$.  If the
inflation took place at the quantum gravity epoch, the above factor is lower still at
$\sim 10^{-120}$.  It is easier to suppose that after inflation was over, the cosmological
constant dropped to zero, rather than try to explain why we see its relic at such a low but highly fine-tuned value.  One may recall that `fine-tuning' was a defect of the standard model, that the idea of inflation was invented to correct. To rationalize fine tuning,
models with $\lambda \propto H^{2}$, $\lambda \propto S^{-2}$ or with  $\lambda \propto t^{-2}$ have been considered by Vishwakarma (2001) to allow for a steady drop of $\lambda$
from its high primordial value to the presently claimed low value.  These models have a somewhat
ad-hoc nature and certainly do not share the elegance of the overall structure of general
relativity.  

\medskip

\noindent  2.  Although it is tempting to place a larger and larger fraction of the total energy
in the $\lambda$ - basket, beyond a certain level the exercise proves counter-productive.  The
stretching of space because of a large $\lambda$ makes gravitational lensing of large redshift objects, a much more frequent phenomenon than
is observationally allowed.  The presently favoured value of $\Lambda_0 \cong 0.7$ is already bursting through this limit, and anything in excess may not be tolerated by the lensing
data (Falco, Kochanek and Munoz, 1998)

\medskip

\noindent 3.  If one uses the `best fit to observations' as a criterion for
estimating $\lambda$, then the angular size-redshift relation for ultra compact radio sources (Jackson and Dodgeson 1997)
implies that $\lambda < 0$.  There is thus a contradiction between two cosmological tests of
the universe going to large redshifts: the $m$-$z$ relation gives $\lambda >0$ whereas the
$\theta$-$z$ relation requires $\lambda<0$ for best-fit model.  Given the uncertainties 
surrounding the cosmological `standard candle' and `standard rulers', we feel that the current
enthusiasm for a positive cosmological constant may be premature.  

\section{An alternative  scenario : the QSSC}

If one is willing to move out of the framework of the standard cosmology, the quasi-steady
state cosmology (QSSC) provides an already existing framework to understand the above 
cosmological issues.  We briefly describe the salient features of this cosmology before
coming to these specific issues.

The QSSC is a non-singular cosmology which arises from a Machian theory of gravity that gives field equations more general than general relativity (Hoyle, Burbidge and Narlikar 1995).  The equations are:

\begin{equation}
R_{ik} -\frac{1}{2} g_{ik} R + \lambda g_{ik} = -8\pi G [T_{ik}^{{\rm matter}}+ T_{ik}^{{\rm creation}}].
\end{equation}

\noindent Here we have taken the speed of light $c=1$ and the first term on the right hand side is the usual energy
momentum tensor of matter, which, in the case of ``dust'' (pressure $p = 0)$ with density $\rho$ and
flow vector $u_i$, becomes

\begin{equation}
T_{ik}^{{\rm matter}} = \rho u_i u_k,
\end{equation}

\noindent while the second term on the right hand side denotes the contribution from a traceless scalar field $C$ of {\it negative} energy and stresses
with gradient $C_l\equiv \partial C/\partial x^l$:

\begin{equation}
T_{ik}^{{\rm creation}} = -f[C_i C_k -\frac{1}{4} C_l C^l g_{ik}].
\end{equation}

The coupling constant $f$ is positive; but with its negative coupling to gravity or spacetime geometry, it generates repulsion and {\it accelerated} expansion of spacetime.  We should point out here that the cosmological constant $\lambda$ in this theory arises from a Machian interaction and {\it its sign is negative}.

For quantitative details of this model, see the paper cited above  (Hoyle, Burbidge and
Narlikar 1995).  General solutions of these equations in the
homogeneous-isotropic case were obtained by Sachs, Narlikar and Hoyle (1996).
The QSSC generally uses the simplest of these, which has a $k = 0$ (flat)
Robertson-Walker spacetime with a scale factor given by

\begin{equation}
S(t) = {\rm exp}(t/P)\times [1 + \eta {\rm~cos}(2\pi \tau(t)/Q)].
\end{equation}

\noindent In this expression the function $\tau(t)$ is very close to $t$, except near the maxima and minima of the scale factor.  The parameters $P$ and $Q$ denote respectively the characteristic time scale of the exponential expansion and the time period of one cycle, while the parameter $\eta$  satisfies the inequality

\begin{equation}
0< \eta <1.
\end{equation}

Evidently for $P >> Q$,  the scale factor describes a universe with a long term exponential expansion together with short term oscillations in which $S(t)$ never reaches the value zero.

The details of this cosmology are found in various papers by Hoyle, Burbidge and Narlikar (1993, 1994 a,b, 1995) as well as in their book ({\it op.cit.,} 2000). The long term expansion in this model is accelerated, as in the original steady state theory.  The repulsive force generated by the  $C-$field is responsible for this feature. However, the $C-$field plays the crucial role of creating new matter {\it without violating any conservation law}.  Although the scale factor describes a smoothed out universe, in actuality the creation is in the form of mini-explosions triggered by the strong gravitational fields of highly collapsed (black-hole like) objects.  It is claimed that the creation phenomenon occurs at various mass scales, ranging from the stellar to the supercluster scales.  

 Within the framework of the QSSC, the cosmic microwave background is explained as the thermalized relic starlight from previous cycles.  The thermal temperature can indeed be worked out in terms of the known stellar activity and its present value can be derived as $\sim 2.7$K. How is thermalization achieved?  The proposed mechanism is via metallic whiskers (Narlikar, et al 1997) which form a component of intergalactic dust. The formation and extinction properties of these whiskers have been discussed in the above paper. Typically the whiskers form from the metallic vapours ejected from supernovae where the metals are synthesized.
Experiments have shown that metallic vapours on cooling, condense into elongated whiskers of $\sim 0.5-1$ mm length and $\sim 10^{-6}$cm cross-sectional radius (Donn and Sears, 1963; Nabarro and Jackson, 1958). It is estimated that a dust density of $\sim 10^{-34}$g cm$^{-3}$ is adequate to achieve the above effect. 
This model can also explain the observed anisotropy of the cosmic microwave background, including the peak at $l \sim 200$ normally referred to as the Doppler peak in standard cosmology (Narlikar, et al 2002). Indeed it is this type of dust that helps explain the magnitude-redshift relation for extragalactic Type Ia supernovae.

\section{The magnitude-redshift relation in the QSSC}

From the above discussion, it is evident that three new aspects are introduced by the 
QSSC vis-a-vis standard cosmology when interpreting the $m$-$z$ relation.  There should
be : \\

\noindent (i)  A negative cosmological constant, \\
\noindent (ii) A repulsive force arising from a negative energy $C$-field,\\
\noindent (iii) Cosmic dust which extinguishes radiation travelling over long distances. 

\medskip

\noindent Of these (i) tends to decelerate the universe, while (ii) tends to accelerate it.  The
effect of (iii), so far as the Type Ia supernova $m$-$z$ test is concerned, is to make the 
distant supernovae progressively dimmer.  The effective magnitude-redshift relation is
given by

\begin{equation}
m(z)={\cal M} + 5 \log (H_0 d_{\rm L})+ \Delta m(z),
\end{equation}
where ${\cal M}\equiv M-5\log H_0+25$, $M$ is the absolute luminosity of the
source and the luminosity distance $d_{\rm L}$ is measured in Mpc.  Here we have
used the notation and definitions of Perlmutter, et al (1999).  The quantity ${\cal M}$
has been referred to by these authors as `Hubble constant free' absolute magnitude. 
The second term on the right hand side of this equation is the standard
luminosity distance $(d_{L})$ term which can be computed for the QSSC.   The product 
$H_{0}d_{L}$ is independent of Hubble's constant.  The last
term denotes the dimming produced by the intergalactic dust and may be
explicitly written as:

\begin{equation}
\Delta m(z) = \frac{\kappa \rho_ {g0}}{H_0}\int^{1+z}_1 \frac{y^2 dy}{(\Lambda_0 + \Omega_0 y^3 + \Omega_{c,0}y^4)^{1/2}},
\end{equation}

\noindent where, $\Lambda_0 = \lambda/3H_0^2$, $\Omega_0 = 8\pi G\rho_0/3H^2_0$
and $\Omega_{c,0} = 8\pi G\rho_{c,0}/3H^2_0$ are the three parameters of this
cosmology corresponding to the cosmological constant, matter density and the
$C-$field energy density. $H_0\equiv 100 ~h_0$ km s$^{-1}$ Mpc$^{-1}$
is the present value of Hubble's constant. The density $\rho_{g0}$ is the
present intergalactic dust grain density in the form of metallic whiskers
and $\kappa$ is the mass absorption coefficient.

Chitre and Narlikar (1976) were the first to discuss the role of intergalactic dust in the
$m$-$z$ relation.  At the time the general belief about the intergalactic medium did not
include dust as a significant component and so the suggestion in the above 
paper that presence of even a small dust-component would enhance the apparent 
magnitudes was largely ignored.  Recently Aquirre (1999) considered the role of dust
in the $m$-$z$ relation, followed by Banerjee, et al (2000).  

A comment is necessary on Eq (7).  In the paper by Banerjee, et al (2000) this formula
had appeared with a typographical error, with the factor $y^2$ missing in the
numerator. The factor was nevertheless included in the actual calculation.   
The calculation redone by Vishwakarma (2002) has slightly lowered 
the value of $\chi^2$ per degree of freedom from $\approx$1.06 to 
$\approx$1.00, thus marginally improving upon the goodness of fit claimed 
originally by Banerjee, et al (2000). 

Here we use this formula to compute the $m$-$z$ curve as predicted by the QSSC.
As reported by Vishwakarma (2002), the calculation is done in two different
ways.
The first is to use the low-redshift measurements to determine the two free
parameters in equation (6), viz. ${\cal M}$ and $\kappa \rho_{g0} h_0^{-1}$,
and then use these values to calculate $\chi^2$ from the high-redshift
measurements including SN 1997ff, the highest redshift supernova observed
so far. Note that the model considered here is completely determined by
$\Lambda_0=-0.358$ and $\eta=0.811$ as considered by Banerjee, et al (2000).
The second method is to use the supernova 1997ff along with the earlier 60
supernovae, taking the high- and low-redshift points together, to get a
best fit curve and then judge the goodness of fit.
If the fit is relatively good, one can argue for the robustness of the
theoretical relation.  
\medskip

\noindent a){\it Method 1}:  Using the earlier work of Banerjee, et al (2000) we carry out the calculation
of magnitudes at specified redshifts, using equation (6).  Six of the sixty supernovae were
dropped from the calculation as the deviations of the observed magnitudes from the 
expected ones are somewhat excessive.  We will comment on this deviation within the
framework of the QSSC later.  For the present we may mention that data on similar
six supernovae had been omitted in obtaining the best-fit models of standard cosmology
by Perlmutter, et al (1999) who had also considered 54 supernovae in their 
primary fit C. 

With the constraint $\rho_{g0}>0$, the sample of 16 low redshift supernovae
gives ${\cal M}\approx 24$ with a wide range  $0<\kappa \rho_{g0} h_0^{-1}<10$
as the best-fitting solution, where $\chi^2$ does not change significantly.
For example, ${\cal M}=24.03$ with $\kappa \rho_{g0} h_0^{-1}=6.5$ gives a
value $\chi^2$ per degree of freedom of 0.8 which represents an excellent fit.
 To fix ideas,
${\cal M}= 24$ gives the corresponding absolute magnitude at peak luminosity of the
supernova as $M = -19.49$ for $h_{0}=0.6$ and $M = -19.16$ for $h_{0}=0.7$.
These
values are in the right region for SN peak magnitudes (Saha, et al 2001).
When we add the new point SN 1997ff ($z=1.755\pm 0.05$, $m=25.68\pm 0.34$)
to the older list of 38 high redshift supernovae with $z\lesssim 0.8$, the
new sample gives a $\chi^2$ per degree of freedom of 1.49 which, though
larger than before, is still acceptable as a reasonable fit.
It would be worthwhile to mention that the supernova 1997ff is claimed
to be on the brighter side due to gravitational lensing effect (Moertell
et al, 2000). If this is the case then an increase in the
measured apparent magnitude of this supernova by only as much as 0.5 improves the
fit considerably by giving
$\chi^2$ per degree of freedom of 1.22. Likewise, a change in the redshift
of this farthest supernova from 1.755 to 1.65 improves the $\chi^2$ per degree
of freedom to 1.33. We demonstrate this to underscore the need for caution in
interpreting data in terms of models till (a) we have data on more supernovae
available and (b) we are more certain about their magnitudes and redshifts.

\bigskip

\noindent b) {\it Method 2} :  Here we include the supernovae 1997ff along with the
54 supernovae from earlier samples in obtaining the best-fit curve.  Naturally the fit
of such a curve to the data will be better than that in the first case.  Here
we report an exercise carried out by one of us (Vishwakarma, 2002) in
which the models are fitted in the QSSC parameter space of
$0>\Lambda_0\geq-0.32$ and  $\rho_{g0}$ in the range
$3-5\times 10^{-34}$g cm$^{-3}$.  The region enclosed by the tilted U-shaped
curve in Figure 1, has the QSSC models with parameters that yield values of
$\chi^2$ per degree of freedom in the range 1.16 to 1.26, that is, values
indicating a satisfactory fit.  Even the model chosen by Banerjee et al (2000),
viz. $\Lambda_{0} = -0.358$ yields a reasonable fit with the $\chi^2$ value
per degree of freedom of 1.27.
Moreover, if we assume that the supernova 1997ff is observed on the brighter
side, as we have mentioned earlier, then an increase in its magnitude by 0.5
improves the $\chi^2$ per degree of freedom to 1.12. Likewise, a change in its
redshift from 1.755 to 1.65 reduces the $\chi^2$ per degree of freedom to 
1.19. 
In Figure 2, we have shown the fitting of this model 
($\Lambda_0=-0.358$, $\eta=0.811$ with $\chi^2$ per degree of freedom $=1.27$)
to the observed data points and compared it with the favoured standard big 
bang model ($\Omega_{\rm total}=1$). It should be noted that the Perlmutter
et al' data taken together with SN 1997ff, do not give the standard flat 
model as the best-fitting model: there the best-fitting model is obtained
with $\Omega_0=0.87$, $\Lambda_0=1.51$ (i.e., $\Omega_{\rm total}=2.38$) with 
a $\chi^2$ value per degree of 
freedom of 1.09. However, if we concentrate our attention on the flat model
($\Omega_{\rm total}\equiv\Omega_0+\Lambda_0=1$),
which is currently favoured, the minimum $\chi^2$ is obtained for the model
$\Omega_0=0.34$ with a $\chi^2$ value per degree of 
freedom of 1.17 (Vishwakarma, 2002).

Refering back to our opening remarks in this paper, it is interesting to compare the old steady state model with the classical Einstein deSitter model
($\Omega_0=1$, $\Lambda_0=0$) for the data of Perlmutter, et al (1999). The 
respective value of $\chi^2$ per degree of freedom for the steady state model is 1.42, compared to the value 1.75 for the Einstein deSitter model. This relative performance may be contrasted with the situation prevailing three decades ago when the $m$-$z$ data on galaxies were claimed to support a {\it decelerating} universe and were used to rule out the steady state model on the grounds that the model predicts an {\it accelerating} universe.

\section{Concluding remarks}

As was originally stated by Banerjee, et al (2000), the adjustable parameter in the above exercise is $\rho_{g0}$, the density of intergalactic dust.  By requiring the fit to be `best' we determine the value of $\rho_{g0}$.  Had the theory been on the wrong track, this value could have come out vastly different from the value needed to thermalize the microwave background.  That it comes out in the correct range must be considered a point in favour of this approach.

It will take us too far from the focus of the present paper to elaborate 
on the details of the microwave background in the QSSC.  In a separate paper
(Narlikar, et al 2002) devoted to the spectrum and fluctuations of the MBR in the
QSSC we have shown that the MBR anisotropy from rich clusters can be revealed by a 
peak in the power spectrum around $l=200$ while smaller peaks are revealed in the
$l \sim 600 - 900$ range from groups of galaxies and perhaps weaker peaks at 
$l \gtrsim 3000$ from individual galaxies or compact groups. 

In the QSSC, the whisker dust is produced and ejected from the vicinity of
supernovae, as discussed by Narlikar et al (1997). Thus it is quite likely 
that dust near the parent supernova may show varying degree of concentration. This can
lead to relatively large fluctuations in the supernova magnitudes, and may
account for the large deviations for the six supernovae omitted from the 
study.

In principle, despite such fluctuations, it is possible to use the $m-z$ test to distinguish between the
standard $\lambda$-based approach and the above QSSC approach.  As we go to higher redshifts, the effect of acceleration as implied by the cosmological constant  will diminish, while that of the intergalactic dust will become  greater. However, great care needs to be exercised before drawing any conclusions from such observations. As the example of the $m-z$ test for galaxies showed in the 1970s, uncertain evolutionary effects tend to dominate at high redshifts. In addition, we have seen that even small changes in the measured values of $m$ or $z$ can greatly alter the conclusion. Finally, as we go to higher redshifts, gravitational lensing may selectively amplify the images and thus push the observed brightness to values higher than normal (Lewis and Ibata 2001). Hence it is prudent to await the accumulation of more type Ia supernovae in the high-$z$ 
basket.

 Geoffrey Burbidge would like to thank one of his collaborators (JVN) for 
hospitality afforded him at IUCAA.

\clearpage

\section*{References}

\noindent Aquirre, A.N. 1999, {\it ApJ}, {\bf 512}, L19

\noindent Bagla, J.S., Padmanabhan, T. and Narlikar, J.V. 1996, {\it  Comm. Astrophys.}, 
{\bf 18}, 275 

\noindent
Banerjee S. K., Narlikar J. V., Wickramasinghe N. C., Hoyle F., Burbidge,
G., 2000, {\it A. J.}, {\bf 119},
\hspace{.5cm} 2583

\noindent Bondi, H. and Gold, T. 1948, {\it M.N.R.A.S.}, {\bf 108}, 252

\noindent Chitre, S.M. and Narlikar, J.V. 1976, {\it AP \& SS}, {\bf 44} 101

\noindent Donn, B. and Sears, G.W. 1963, {\it Science}, {\bf 140}, 1208

\noindent Falco, E.E., Kochanek, C.S. and Munoz, J.A. 1998, {\it Ap. J.}, {\bf 494}, 47

\noindent Filippenko, A.V. 2001, {\it P.A.S.P. Millennium Essay}, {\bf 113}, 1441

\noindent Friedmann, A. 1922, {\it Z.Phys.}, {\bf 10}, 377

\noindent Hoyle, F. 1948, {\it M.N.R.A.S.}, {\bf 108}, 372

\noindent Hoyle, F., Burbidge, G. and Narlikar, J.V. 1993, {\it Ap. J.}, {\bf 410}, 437

\noindent Hoyle, F., Burbidge, G. and Narlikar, J.V. 1994$a$, {\it M.N.R.A.S.}, {\bf 267}, 1007

\noindent Hoyle, F., Burbidge, G. and Narlikar, J.V. 1994$b$, {\it A\& A}, {\bf 289}, 729

\noindent Hoyle, F., Burbidge, G. and Narlikar, J.V. 1995, {\it Proc. Roy. Soc. A.}, {\bf 448}, 191

\noindent Hoyle, F., Burbidge, G. and Narlikar, J.V. 2000, {\it A Different Approach to Cosmology},
Cambridge,
\hspace{.5cm} p. 201

\noindent Hubble, E. 1929, {\it Proc. Nat. Acad. Sci.}, {\bf 15}, 168

\noindent Jackson, J.C. and Dodgson, M. 1997, {\it M.N.R.A.S.}, {\bf 285}, 806

\noindent Lemaitre, G. 1927, {\it Ann. de la Societe Scientifique de Bruxelles}, {\bf 47}, 49

\noindent Lewis, G.F. and Ibata, R.A. 2001, {\it preprint} : astro-ph/0104254 v2

\noindent Moertsell E., Gunnarsson C., Goobar A., astro-ph/0105355

\noindent Nabarro, F.R.N. and Jackson, P.J. 1958, in {\it Growth and Perfection in Crystals}, 
eds. R.H. Duramus, et al, (J. Wiley, New York)

\noindent Narlikar J. V., Wickramasinghe N. C., Sachs R., Hoyle F., 1997, Int. J. Mod.
Phys. D, 6, 125

\noindent Narlikar, J.V. and Padmanabhan, T. 2001, {\it Ann. Rev. A \& A.}, {\bf 39}, 211

\noindent Narlikar, J.V., Vishwakarma, R.G., Hajian A., Souradeep T.,
Burbidge, G., Hoyle F., 2002, {\it preprint}

\noindent Perlmutter, S. et al, 1999, {\it Ap.J.}, {\bf 517}, 565

\noindent Riess, A.  et al. 1998, {\it A.J.}, {\bf 116}, 1009

\noindent Riess, A. et al. 2000, {\it Ap.J.}, {\bf 536}, 62

\noindent Riess, A. et al. 2001, {\it Ap.J.}, {\bf 560}, 49

\noindent Sachs R., Narlikar J. V., Hoyle F., 1996, A\&A, 313, 703

\noindent Saha A. et al. 2001, {\it Ap.J}, {\bf 562}, 314

\noindent Vishwakarma, R.G.  2001, {\it Class. Qu. Grav.}, {\bf 18}, 1159

\noindent Vishwakarma, R.G. 2002, {\it M.N.R.A.S.}, {\bf 331}, 776

\noindent Weinberg, S. 1989, {\it Rev. Mod. Phys.}, {\bf 61}, 1

\begin{figure}
\resizebox{\textwidth}{!}{\includegraphics{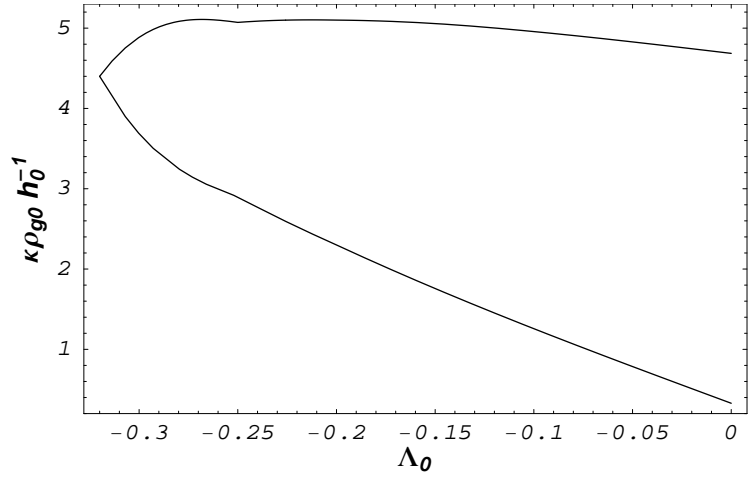}}
\caption{The tilted U-shaped contour shows a parameter-space of interest
in the QSSC. The points of the
contour have $\chi^2$ per degree of freedom in the range $1.16 - 1.26$.
The coupling constant $\kappa$ and
the present density of the whisker grains $\rho_{g0}$
are measured, respectively,
in units of $10^5$ cm$^2$ g$^{-1}$ and $10^{-34}$ g cm$^{-3}$.
[Adapted from Figure 6 of Vishwakarma (2002)]}
\end{figure}

\begin{figure}
\resizebox{\textwidth}{!}{\includegraphics{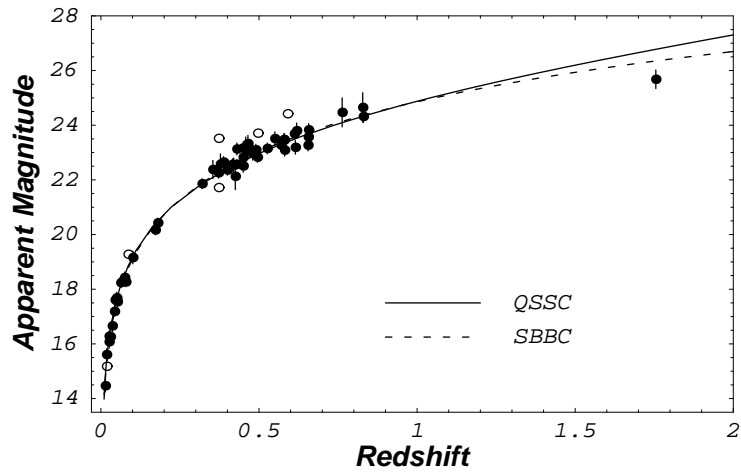}}
\caption{Hubble diagram for 55 supernovae is shown. The theoretical curves
represent the best-fitting flat QSSC ($\Lambda_{0} = -0.358$, $\eta=0.811$) 
and flat SBBC ($\Omega_0=0.34$) models.  The six omitted points are shown
by open circles.}
\end{figure}

\end{document}